# Modeling of durability of polyelectrolyte membrane of O$_2$/H$_2$ fuel cell

**Vadim V. Atrazhev**[1,2] and **Sergei F. Burlatsky**[3]

[1] Russian Academy of Science, Institute of Biochemical Physics, Kosygin str. 4, 119334, Moscow, Russia
[2] Science for Technology LLC, Leninskiy pr-t 95, 119313, Moscow, Russia
[3] United Technologies Research Center, 411 Silver Lane, East Hartford, CT 06108, USA

## 1. Introduction

In this paper, we discuss critical aspects of the mechanisms and features of polymer proton exchange membrane (PEM) degradation in low-temperature H$_2$/O$_2$ fuel cell. The low-temperature fuel cell with PEM is a major candidate to replace internal combustion engine in vehicles. The membrane is a critical component of the fuel cell that conducts protons while preventing the flow of electrons through it. Additionally, the membrane separates the fuel and the oxidant streams. Membrane degradation continues to present challenges for PEM fuel cells to meet the required lifetimes in transportation applications. This is especially the case when the operating conditions favor higher temperatures and lower RH to reduce the fuel cell system balance of plant cost, weight, and volume. Under these conditions, the chemical attack of the membrane by oxidizing radicals is highly accelerated [1,2].

High durability of the PEM, often 5–20 khr, is required for many applications. Understanding of degradation mechanisms and accurate lifetime predictions are critical for the commercial viability of these applications because such long times are not usually available to assure the lifetime of the cell-stack design before fielding significant numbers of commercial units. Both chemical and mechanical factor are important for degradation of perfluorosulfonic acid (PFSA)-based polymer membranes (e.g., Nafion). Chemical degradation involves two major steps: radical generation and radical attack that is likely to proceed at the linear polymer backbone [3,4]. This attack is enhanced if the backbone is not completely fluorinated, particularly if it contains a carboxylic end-group. As first discussed in [3] the polymer backbone proceeds to degrade by the following general mechanism

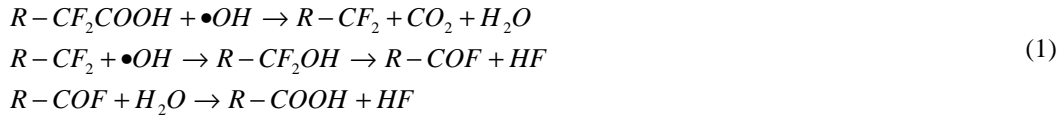
$$\begin{aligned}R - CF_2COOH + \bullet OH &\to R - CF_2 + CO_2 + H_2O \\ R - CF_2 + \bullet OH &\to R - CF_2OH \to R - COF + HF \\ R - COF + H_2O &\to R - COOH + HF\end{aligned} \qquad (1)$$

In this paper, we focused on chemical mechanism of OH radical generation and their distribution in operational fuel cell. According to the current concept, free radicals are generated from hydrogen and oxygen crossover gases at the surface of Pt particles that precipitated in the membrane. We explicitly calculate Pt precipitation rate and electrochemical potential distribution in the membrane that controls it. Based on radical generation rate and Pt distribution we calculate degradation rate of the membrane taking advantage of simple kinetics equations. However, the chemical details of membrane attack by radicals and mechanical degradation are out of the scope of this paper. The paper is organized as follows. The model of Pt precipitation in the membrane is presented in section 2. The chemical aspects of free OH radicals generation in the membrane are discussed in section 3.1. The macroscopic model of the membrane degradation is presented in section 3.2. Conclusion from the modeling results are presented in section 4.

## 2. Pt precipitation in membrane

The chemical degradation of PEM in fuel cell is catalyzed by Pt particles precipitated in the membrane during operation. Pt is a commonly used catalyst in the fuel cell electrodes. Instability of Pt at high potentials is greater and is aggravated by potential cycling [5]. The Pt ions dissolved at the cathode, diffuse through the membrane towards the anode and precipitate in the membrane forming narrow band of Pt particles. The location of the Pt band in the membrane is governed by potential of Pt particles in the membrane. This potential is controlled by relatively small crossover diffusion flux of oxygen from the cathode side and hydrogen from the anode side. The crossover gases react at Pt particles surface. Competition of oxygen reduction reaction (ORR) and hydrogen oxidation reaction (HOR) governs the local potential of Pt particles in the membrane [6]. This potential is high near the cathode and low near anode. It abruptly drops in specific point $X_0$ in the membrane bulk. The Pt ions dissolved in the membrane are reduced to Pt atoms in vicinity of $X_0$ and then are agglomerated into Pt particles that form Pt band [7].

Pt electrochemically dissolves in the cathode by reaction

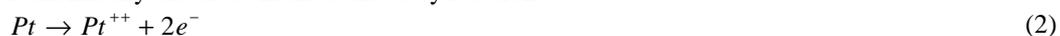
$$Pt \to Pt^{++} + 2e^- \qquad (2)$$

The Pt ions diffuse into the membrane driven by the concentration gradient. Hydrogen crossover from the anode reduces Pt ions forming Pt atoms by reaction

$$Pt^{++} + H_2 \rightarrow Pt + 2H^+ \tag{3}$$

The Pt atoms diffuse and agglomerate to form Pt particles, $Pt_n$, to minimize their high surface energy.

$$Pt_n + Pt \rightarrow Pt_{n+1} \tag{4}$$

Pt ions also precipitate electrochemically at low potentials on the Pt particles through reaction (5a). Subsequent precipitation would increase the Pt particle potential and stop precipitation. The Pt particle potential is also increased by reduction of oxygen crosser by reaction (5b). However, the potential of platinum particles is simultaneously decreased by $H_2$ crossover oxidation through reaction (5c) that supports further Pt precipitation.

$$Pt_n + Pt^{++} \rightarrow Pt_{n+1}^{++} \tag{5a}$$

$$Pt_n + O_2 + 4H^{++} \rightarrow Pt_n^{4+} + 2H_2O \tag{5b}$$

$$Pt_{n+1}^{++} + H_2 \rightarrow Pt_{n+1} + 2H^+ \tag{5c}$$

The flux of Pt ions, hydrogen and oxygen crossover gases towards Pt particle surface increases with Pt particle size. The dynamics of platinum precipitation and growth is governed by the local potential and size of the Pt particles that is calculated in the next section.

### 2.1 Potential of Pt particles in membrane

Potential of Pt particles in the membrane is governed by kinetics of oxygen reduction reaction (ORR) and hydrogen oxidation reaction (HOR) at the surface of Pt particles in the membrane. Equilibrium mixed potential resulting from parallel reactions of oxidation and reduction is well understood. For typical relatively slow reactions, the equilibrium mixed potential value lies between the equilibrium potential of ORR and HOR. At the cathode, the equilibrium mixed potential is equal to open circuit cathode potential (~980 mV) and abruptly drop to approximately 500 mV at membrane/cathode interface as shown by the red line in Fig. 1. In the membrane bulk, the equilibrium mixed potential weakly changes and is equal to approximately 300 mV at membrane/anode interface. Then it abruptly drops at close proximity to the anode to open circuit anode potential (~0 mV). Since the equilibrium mixed potential is much lower than the equilibrium potential of reaction (2) (1.188 V) the Pt band would form at the cathode/membrane interface. That contradicts to experimental observations from [8,9], where Pt band was observed in the membrane bulk closer to the cathode/membrane interface. Moreover, Liu and Zuckerbrod in [10] experimentally discovered unexpected behavior of the potential of catalyst microelectrode probe inserted into membrane that was incorporated into $H_2/O_2$ fuel cell. As expected, the measured potential of the probe is close to equilibrium potential of ORR (1.23 V) near cathode and close to equilibrium potential of HOR (0 V) near anode. However, the potential undergoes abrupt change in the membrane bulk, as shown in Fig. 1. The location of the potential abrupt change is denoted as $X_0$.

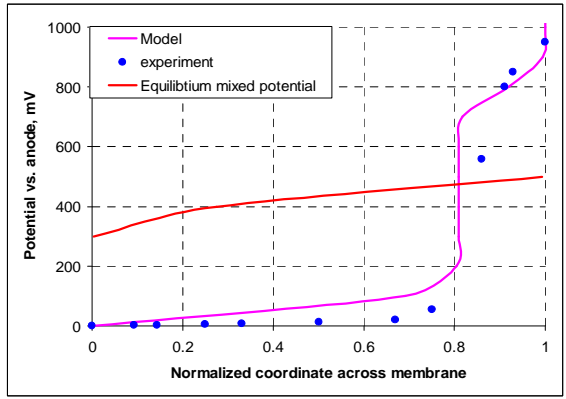

Fig. 1 Experimentally measured [10] potential of Pt probe in PEM as a function of normalized coordinate across membrane (dots). Model prediction and equilibrium mixed potential are shown in the plot by lines.

Unexpected dependence of the particle potential on coordinate across the membrane is caused by the local transport limitations near the particle [6]. The Pt particle potential is governed by the balance of the currents of ORR, $j_{ORR}$, and HOR, $j_{HOR}$, at the particle surface

$$i_{HOR}(C_{H_2}(x), \Phi) + i_{ORR}(C_{O_2}(x), \Phi) = 0 \tag{6}$$

Here $C_{H2}$ and $C_{O2}$ are hydrogen and oxygen concentrations at the surface of Pt particle, $\Phi$ is Pt particle potential and $x$ is dimensionless coordinate across membrane. The rate of heterogeneous electrochemical reaction depends on the local concentration of reagents at the surface of the catalyst and local potential of the catalyst through Butler-Volmer equation

$$j(C, \Phi) = j_0 \left( C \exp\left( \frac{\alpha F(\Phi^0 - \Phi)}{nRT} \right) - \exp\left( -\frac{(1-\alpha)F(\Phi^0 - \Phi)}{nRT} \right) \right) \tag{7}$$

Here $\Phi^0$ is equilibrium potential of reaction (1.23 V for ORR and 0 V for HOR), $j_0$ is exchange current density, $C$ is reagent concentration at the catalyst surface, $n$ is electron transfer number ($C=C_{H2}$ and $n=2$ for hydrogen; $C=C_{O2}$ and $n=4$ for oxygen).

Local transport limitations reduce the concentrations of the reagents (H$_2$ and O$_2$) at the Pt particle surface as compared with concentration in the membrane bulk[1]. Here we model the Pt particle as a sphere with the radius $a$. The typical size of Pt particles detected in PEM is 5–50 nm as reported in [11], which is by 3 orders of magnitude smaller than the membrane thickness, $L$. That justifies applicability of scale separation approach. We calculate the concentration of each reagent, oxygen or hydrogen, at the surface of Pt particle through the concentration of reagent in the membrane bulk, $C_b$, and reaction rate, $j$:

$$C(x) = C_b(x) - \frac{j(C(x), \Phi)}{j_{\lim}} \quad (8)$$

Here $j_{lim}$ is a local limiting current of the reagent transport to Pt particle, $x$ is a coordinate across the membrane. It is a slow varying variable. The limiting current of the reagent to Pt particle in the membrane is calculated by equation: $j_{\lim} = nFC_{ref}HD/a$. Here $H$ is dimensionless solubility of gas in the membrane, $D$ is gas diffusion coefficient in the membrane ($D=D_{H2}$ for hydrogen and $D=D_{O2}$ for oxygen), $F$ is Faraday constant, $C_{ref}$ is reference concentration of gas. Substituting equation (8) into (7) and solving obtained equations with respect to $j_{ORR}$ and $j_{HOR}$ we obtain the equations for the rates of ORR and HOR with local transport limitations to individual Pt particle. These equations relate the rates of ORR and HOR with the local potential of the particles and gases concentrations in the membrane bulk, $C_b(x)$:

$$j(C_b(x), \Phi) = j_0 \frac{\left(C_b(x)\exp\left(\frac{\alpha F(\Phi^0 - \Phi)}{nRT}\right) - \exp\left(-\frac{(1-\alpha)F(\Phi^0 - \Phi)}{nRT}\right)\right)}{\left(1 - \frac{j_0}{j_{\lim}}\exp\left(\frac{\alpha F(\Phi^0 - \Phi)}{nRT}\right)\right)} \quad (9)$$

Substituting equation (9) for ORR and HOR into charge balance equation (6) we obtain equation that relates the Pt particle potential to the local concentrations of the dissolved gases (oxygen and hydrogen) in the membrane bulk:

$$i_{HOR}(C_{H_2,b}(x), \Phi) + i_{ORR}(C_{O_2,b}(x), \Phi) = 0 \quad (10)$$

The disturbance of mesoscopic concentrations profile, $C(x)$, by the probe particle is negligible if $L \gg a$. Oxygen concentration at cathode/membrane interface is determined by oxygen concentration in the cathode gas and is equal to zero at anode/membrane interface due to fast oxygen consumption at Pt catalyst at the anode potential. Hydrogen concentration at anode/membrane interface is determined by hydrogen concentration in the anode gas and is equal to zero at cathode/membrane interface due to fast hydrogen consumption at Pt catalyst at the cathode potential. In the bulk of the membrane, the concentrations of the gases are linear function of dimensionless coordinate, $x$, across membrane

$$C_{O_2,b}(x) = C^0_{O_2} x, \qquad C_{H_2,b}(x) = C^0_{H_2}(1-x) \quad (11)$$

Substituting equations (11) into (10) we obtained the final equation for potential of probe Pt particle in the membrane. Numerical solution of this equation is shown in Fig. 1 by the pink line. The $X_0$ value is calculated analytically, $x_0 = H_{H_2}D_{H_2}C^0_{H_2} / \left(H_{H_2}D_{H_2}C^0_{H_2} + 2 \cdot H_{O_2}D_{O_2}C^0_{O_2}\right)$, where H is gas solubility in the membrane and D gas diffusion coefficient the membrane.

Qualitative explanation of abrupt change of potential near $X_0$ follows from the theory of diffusion-controlled reactions. In the region between $X_0$ and the cathode, the HOR is limited by hydrogen diffusion. In this region, the hydrogen flux approaches the limiting diffusion flux of hydrogen. The hydrogen concentration at the probe surface approaches zero. To maintain zero total current through the probe, the oxygen reduction rate should be equal to the half of hydrogen oxidation rate. However, the oxygen flux is smaller than the oxygen limiting flux, the oxygen concentration is finite (non-zero) and the probe potential is determined by the ORR. The potential of the probe between $X_0$ and the anode is approximately equal to the equilibrium potential of oxygen reduction reaction. Similar consideration is valid for the region between $X_0$ and the anode. In this region, electrochemical reactions are limited by oxygen diffusion, oxygen concentration at the probe surface is close to zero, and the potential of the probe drops towards the equilibrium potential of hydrogen oxidation reaction. The distribution of precipitated Pt in the narrow band in the membrane is caused by the singular dependence of Pt particle potential in the membrane on coordinate across the membrane.

2.2 The model of Pt precipitation in membrane

At the nucleation stage, the size of Pt nuclei is small and electrochemical reactions are suppressed. Electrochemical reactions are not energetically favorable on a small Pt particle, containing just a few Pt atoms, because transfer of one electron leads to high charge density on the Pt particle. During this stage, the platinum ions from the cathode react with hydrogen forming atomic Pt, as indicated by equation (3). Atomic Pt agglomerates making Pt particles, as indicated by equation (4). At this stage, the hydrogen concentration in the membrane is much higher than the Pt particle concentration. Therefore, reaction (3) dominates reaction (5a). Reaction (3) is very fast (diffusion controlled). Therefore, most Pt ions are reduced by hydrogen on the cathode side of the membrane and only a very small fraction can migrate to the anode. The aggregation process of the small Pt particles through reaction (4) is also fast with diffusion-controlled rate constant, $K_{agg} \approx 16\pi D_{Pt} a_{Pt} \approx 10^{-12} \, cm^3/s$. To calculate the final concentration of Pt nucleus at the end of nucleation stage we utilized the theory of diffusion-controlled agglomeration [12]. According to [12], the total concentration of Pt particles in the membrane at nucleation stage is

$$n_{Pt}(t) = n_{Pt}^0 \tanh(t/\tau_{agg}) \quad (12)$$

---

[1] The concentration in the bulk is equal the concentration at the distance from the particle that is much larger than particle size

Here $n_{Pt}^o = \sqrt{2J/K_{agg}}$, $\tau_{agg} = \sqrt{2/JK_{agg}}$ and $J \approx 10^{13} cm^{-3} s^{-1}$. Here, $n_{Pt}^o \approx 10^{13} cm^{-3}$ is characteristic Pt particle concentration and $\tau_{agg}$~1s is characteristic aggregation time.

The stage of nucleus growth starts when the mean size of Pt particles exceeds the critical radius $r_{c,el}$~2.5nm and the electrochemical reactions (ORR, HOR and Pt presipitation) on the Pt particle surfaces switch on. That effects the distribution of gases in the membrane due to ORR and HOR at the Pt particles. Simultaneously, Pt particles become immobile when they reach the mean size of the water clusters in hydrated Nafion membrane, $r_c$~2.5nm. The growth of Pt particles occurs as follows. The Pt ions freely diffuse from the cathode to the plane $X_0$, where they turn into atomic platinum due to hydrogen excess. These neutral Pt atoms aggregate with the Pt particles resulting in maximum growth of Pt particles near $X_0$.

A one-dimensional model along the thickness of the membrane was developed to capture the mechanism of platinum dissolution, precipitation, and growth within the membrane [7]. The reaction-diffusion equations for $O_2$, $H_2$, Pt atoms, Pt ions and Pt particles were derived and solved numerically and the Pt particle size distribution in the membrane, $r(x,t)$, was calculated. The model equations and boundary conditions are summarized in Table 1.

**Table 1** Model equations and boundary conditions.

| Notation | Description | Equation | Boundary conditions |
|---|---|---|---|
| $C_{O_2}(x,t)$ | $O_2$ concentration, mol/cm$^3$ | $D_{O_2} \frac{\partial^2 C_{O_2}}{\partial x^2} = K_{O_2}$ | $C_{O_2}\vert_{AMI} = 0 \quad C_{O_2}\vert_{CMI} = H_{O_2} C_{O_2}^{chan}$ |
| $C_{H_2}(x,t)$ | $H_2$ concentration, mol/cm$^3$ | $D_{H_2} \frac{\partial^2 C_{H_2}}{\partial x^2} = K_{H_2} + K_{Pt^{++}H_2}$ | $C_{H_2}\vert_{CMI} = 0 \quad C_{H_2}\vert_{AMI} = H_{H_2} C_{H_2}^{chan}$ |
| $C_{Pt^{++}}(x,t)$ | Concentration of $Pt^{++}$ ions in solution, mol/cm$^3$ | $D_{Pt^{++}} \frac{\partial^2 C_{Pt^{++}}}{\partial x^2} = -K_{Pt^{++}} + K_{Pt^{++}H_2}$ | $-D_{Pt^{++}} \frac{dC_{Pt^{++}}}{dx}\vert_{CMI} = J_{Pt,total} \quad C_{Pt^{++}}\vert_{AMI} = 0$ |
| $C_{Pt}(x,t)$ | Concentration of free Pt atoms in solution, mol/cm$^3$ | $D_{Pt} \frac{\partial^2 C_{Pt}}{\partial x^2} = -K_{Pt^{++}H_2} + K_{Pt}$ | $C_{Pt}\vert_{CMI} = 0 \quad C_{Pt}\vert_{AMI} = 0$ |
| $\Phi(x,t)$ | Electric potential of Pt particle, V | $2K_{O_2} - K_{H_2} - K_{Pt^{++}} = 0$ | |
| $r(x,t)$ | Radius of Pt particles, cm | $\frac{4\pi r^2 n_{Pt}}{v_{Pt}} \frac{\partial r(x,t)}{\partial t} = K_{Pt} - K_{Pt^{++}}$ | Initial condition r(x,0) = 3 nm |

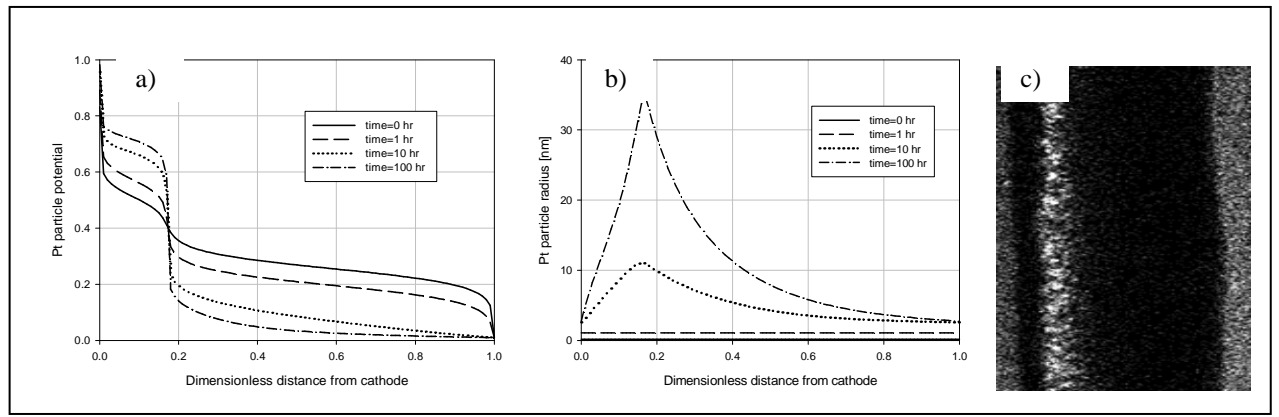

**Fig. 2** (a) calculated evolution of Pt particle potential in membrane, (b) calculated evolution of Pt particle radius in the membrane, (c) SEM image of the membrane electrode assembly after potential cycling under H2/air conditions **[8]**.

The model predictions for platinum evolution in the membrane under potential cycling conditions are shown in Fig. 2. The evolution of potential distribution of Pt particles in the membrane is shown in Fig. 2(a). The potential drop in $X_0$ becomes more pronounced with time because diffusion limitation becomes stronger (limiting current density at Pt surface decreases) with increase of the Pt particles size. The evolution of Pt particle radius in the membrane is shown in Fig. 2(b). For comparison, SEM image of the membrane electrode assembly after potential cycling under $H_2$/air conditions from [8] is presented in Fig. 2(c). Comparison of SEM image and model prediction indicates that the model correctly predicts location of Pt band in the membrane.

## 3. Chemical degradation of the membrane

The mechanism of free radicals formation in the membrane is discussed in this section. Crossovers of oxygen and hydrogen in the membrane react at the Pt precipitated in the membrane. At high potential, at the cathode, catalytic oxygen reduction reaction (ORR) proceeds with $H_2O$ formation, while ORR involves $H_2O_2$ formation as by product at low potential, at the anode, [13]. Direct formation of OH radicals by ORR is possible at intermediate potentials [14]. As indicated in Fig. 1, the intermediate potential are predicted and observed at the location of Pt band in the membrane. Therefore, oxygen crossover is reduced in the membrane with generation of free OH radicals. The radicals attack the polymer molecules of the membrane causing chemical degradation. Intensive chemical degradation leads to pinholes/cracks formation at $X_0$ (Pt band location) in the membrane [7,15]. Chemical degradation also decreases the membrane mechanical strength, that leads to accelerated mechanical failure.

### 3.1 Mechanism of free radicals formation in fuel cell

Conventional understanding of mechanism of polymer membrane chemical degradation in fuel cell involves the hydrogen peroxide as intermediate product of OH radical formation in the membrane. Oxygen crossover diffuses through the membrane to the anode and is reduced here with formation of $H_2O_2$. This peroxide diffuses into the membrane and reacts with contaminants (metal ions) with OH radical formation. However, the estimate for OH generation rate by this mechanism is much lower than the membrane degradation rate measured by the fluoride emission rate from operational fuel cell [15]. That indicates that conventional mechanism of OH radical generation through decomposition of $H_2O_2$ might be not a leading mechanism of free radicals generation in PEM.

Another mechanism of OH radicals generation in the PEM fuel cell was identified by quantum chemical study of ORR at Pt [14]. Density functional theory calculations were performed using the B3LYP hybrid density functional theory. The methods outlined in Sidik and Anderson [16] were used to determine the effects of electrode potential on the transition state activation energy.

The energetic of various ORR intermediates were calculated for relatively small Pt ensembles. The results indicate that formation of $H_2O$ is kinetically favored over the 0–1 V potential range for $O_2$ adsorbed at two Pt atoms. In agreement with results of [16], the electro-reduction of adsorbed $O_2$ is favored over the dissociation of $O_2$ with the 1st electron transfer being rate limiting. The desorption of any $OH_{ad}$ from the Pt surface is kinetically unfavorable with activation energy $E_a$=2.38 eV. Oxygen competes for Pt sites (atoms) with adsorbed $OH_{ads}$ and with of other adsorbed intermediates, such as $Cl^-$ ions if present [17] or CO in contaminated hydrogen or reformate-based systems. Such species will disrupt the continuous Pt ensembles available to the ORR and favor the production of hydrogen peroxide [18] detected by rotating ring-disk electrode (RRDE) measurements.

To understand the reaction path and calculate the barriers for ORR at a poisoned Pt surface we modeled $O_2$ adsorbed at a single Pt site through a single oxygen atom. The mechanism of ORR at the poisoned Pt surface is illustrated in Fig. 3. It involves a single bound OOH that can either reduce further to adsorbed peroxide, or dissociate into a free OH radical. The latter process is non-electrochemical and its rate does not explicitly depends on the potential. However, at low potential it is suppressed by competition with electrochemical formation of adsorbed $H_2O_2$. The further reduction to adsorbed peroxide is strongly favored at 0.0 V, but less so at 1.0 V; therefore, the rate-determining step in peroxide generation may be the 2nd reduction step. $O_2$ reduction to hydrogen peroxide requires significant overpotential and occurs only at E<0.7 V. Adsorbed peroxide can desorb or decompose with generation of free OH radical. According to our calculations, there is no barrier for the latter step, so it is likely the viable path for direct radical generation.

To summarize, the direct generation of free OH radicals is expected on Pt surfaces under conditions where oxygen reduction is hindered by other adsorbed species. At bare Pt (i.e. oxygen molecule bonds with two Pt sites) oxygen is reduced to water only at any potential without $H_2O_2$ or OH radical generation. At bare Pt, peroxide also reduces to water at any potential without OH generation. However, OH radicals are generated directly by oxygen reduction reaction at high potential if oxygen molecule bonds with one Pt site (neighbor sites are blocked by some adsorbed species). Peroxide also decomposes with formation of OH radicals at any potential if bonds with one Pt site. Free OH radicals are trapped rapidly (w/o energy barrier) at Pt at any potential, which significantly reduces the selectivity for radical attack on adjacent polymer.

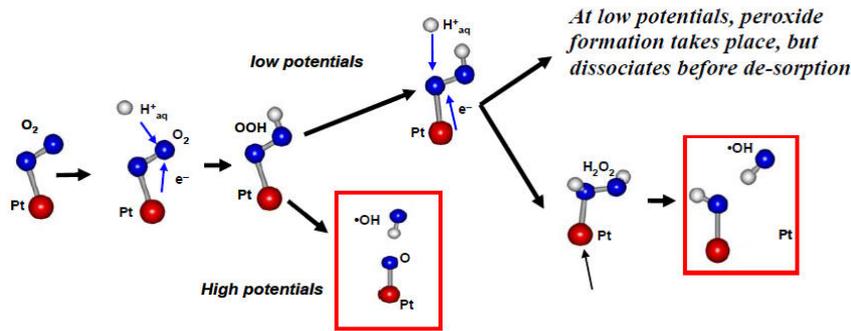

**Fig. 3** Direct radical-formation mechanism schemes at low and high potentials

3.2 The chemical degradation model

A three-layer 1D model, including the anode, cathode, and membrane, was developed to describe the chemical degradation of the membrane. The model is based on the following models/assumptions. (i) The Pt particles distribution in the membrane was calculated by the model of Pt precipitation (see section 2.2) (ii) The ORR at the Pt particles is the source of hydroxyl (OH) radical, water and hydrogen peroxide. The potential of the particles was calculated as a function of local oxygen and hydrogen concentrations. The potential-dependent rates of the oxygen reduction to water and hydrogen peroxide were determined from published experimental data [13]. The oxygen reduction selectivity to radical formation is a fitting parameter in the model. (iii) Hydroxyl radicals can react with Nafion and generate hydrogen fluoride (HF). The rate constant of this reaction is also a fitting parameter. According to [1,3], one moles of radicals generates two moles of hydrogen fluoride. (iv) The Pt particles or carbon also quench radicals. Ab initio results indicate that hydroxyl radical reactions with these surfaces are diffusion limited [14]. (v) Peroxide can also be generated at carbon-supported Pt surfaces with rate constants based on literature studies [3]. It is decomposed on Pt surfaces with rates obtained from RRDE measurements. The selectivity of this decomposition to radicals is the third and the last fitting parameter. (vi) The potentials of isolated Pt particles in the membrane are determined by the local hydrogen/oxygen concentrations and calculated by approach presented in section 2.1.

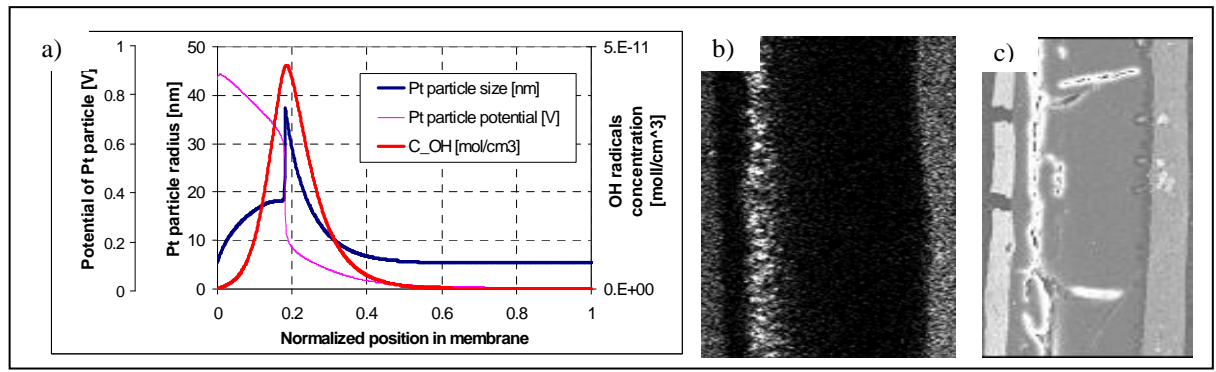

**Fig. 4 (a)** The model prediction for Pt particles size, potential and OH radical concentration as a function of normalized coordinate across the membrane, (b) SEM image of the membrane electrode assembly after potential cycling under H2/air conditions **[8]**, (c) SEM image of degraded membrane after potential cycling in $H_2$/air conditions.

The conservation reaction-diffusion equations for the six species ($O_2$, $H_2$, $H_2O$, $H_2O_2$, OH, HF), distributed in the membrane through-plane spatial dimension, are solved for the three layers of the MEA to estimate the FER. The model calibration was performed through experimental data on FER from fuel cell measured by ion chromatography in anode and cathode gas condensate [1]. The model prediction for Pt particles size, potential and OH radical concentration as a function of normalized coordinate across membrane are shown in Fig. 4(a). The SEM images of Pt precipitated in the membrane (Fig. 4(b)) and degraded membrane (Fig. 4(c)) after potential cycling in $H_2$/air conditions are also presented in the same figure for comparison with model prediction. The OH radicals are distributed in the membrane in narrow band near $X_0$=0.2 in $H_2$/air conditions, as is observed in SEM image and predicted by the model. Maximum of OH radical concentration coincides with maximum of Pt particle distribution (maximal particle size) in the membrane.

Hydrogen peroxide decomposition rate was calculated to understand the hydrogen peroxide mechanism contribution to membrane degradation. According to [1], no Fenton reagents (contaminating metal ions) were detected in either the condensate water samples or in the electron microprobe analysis. Therefore, we assume that the Pt particles throughout the membrane thickness provide the only catalyst for radical generation in the membrane. The simulations were performed for various membrane thicknesses to observe the corresponding magnitude of FERs, using the measured $H_2O_2$ decomposition rate as the input. The results are shown in Fig. 5. The average $H_2O_2$ concentration in the membrane varied in the interval from 100 to 200 ppm. To reconcile these low concentration with experimentally observed FER levels, one would have to assume unrealistically high rates of peroxide decomposition that are by 3 orders of magnitude higher than measured even if unrealistically high 100% yield of peroxide to radicals is assumed. That implies that peroxide decomposition might be not the leading mechanism for free radicals generation in PEM.

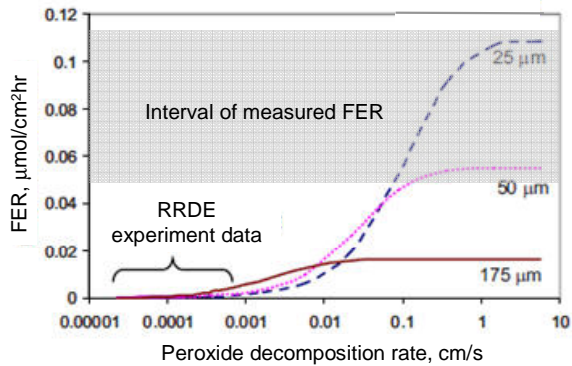

**Fig. 5** FER vs. peroxide decomposition rate and membrane thickness assuming that the peroxide (indirect) mechanism only is responsible for radical generation.

The model predicts dependence of FER on key factors that impact membrane degradation: amount of platinum, membrane thickness, the hydration level of the membrane, the concentration of the crossover reactants, and the temperature. Unexpected dependences of FER on Pt particles size and membrane thickness are predicted by the model, as shown in Fig. 6. For low Pt loading in the membrane (small spacing between particles) dependences of FER on the Pt particle size and the membrane thickness have a maximum. That is due to twofold role of Pt in the membrane. On the one hand, the Pt particles are sources for free OH radicals in the membrane. On the other hand, the Pt particles are sinks for OH radicals due to diffusion-controlled quenching of the radicals at Pt surface. Competition of these two processes results in non-monotonic dependence of membrane degradation rate on Pt loading. The low loading of small Pt particles in membrane causes fast degradation of the membrane. In contrast, the high loading of large Pt particles in the membrane mitigates membrane from chemical degradation, as was reported in [9,19].

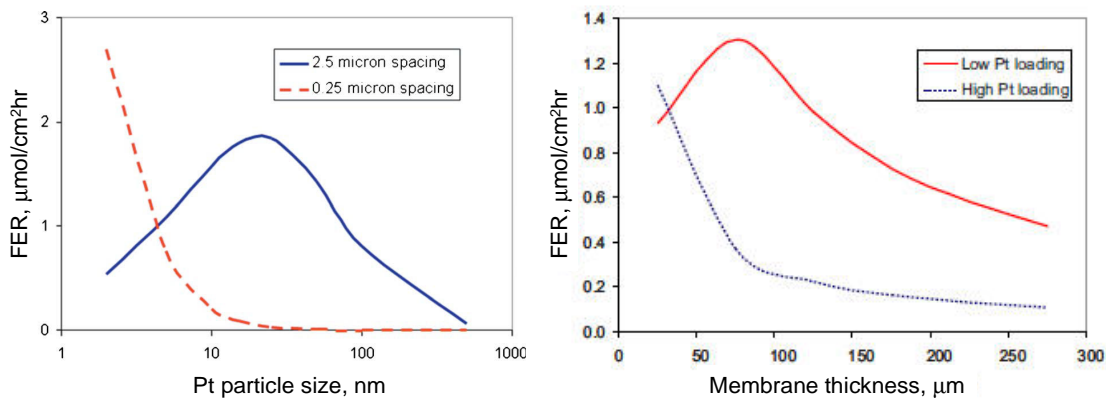

**Fig. 6** The chemical degradation model prediction for FER as a function of (a) Pt particle size and (b) membrane thickness

## 4. Conclusions

A physics-based model that predicts the location and rate of chemical degradation of PEM fuel cell membrane was developed. The model includes three sub-models: (i) the model of Pt precipitation in the membrane and (ii) atomistic model of OH generation/decomposition pathways in the membrane and (iii) the model of chemical degradation. The model of Pt precipitation predicts Pt distribution in narrow band in the membrane near specific point $X_0$. Location of $X_o$ depends on $O_2/H_2$ concentrations in the electrodes and motilities in the membrane. The atomistic model of OH radical generation indicates that free OH radicals and hydrogen peroxide ($H_2O_2$) are generated as by-products of ORR at Pt surface contaminated by absorbed species (CO, $SO_3^-$). Hydrogen peroxide can be also decomposed with formation of free OH radicals under the same conditions. Thus, free OH radicals are generated at Pt particles in the membrane directly as by-product of ORR or indirectly through $H_2O_2$ decomposition. The model predicts that direct mechanism is at least by three orders of magnitude more effective then indirect mechanism. Moreover, the chemical degradation model prediction agrees with experimentally measured fluoride emission rate for direct mechanism and is by three orders of magnitude lower for indirect mechanism. That implies that the direct mechanism is prevailing in PEM. The model of chemical degradation provides a consistent picture of membrane decay modes and the contributing factors that interact in a complex way. The model predicts and reconciles several published literature experimental trends. The amount of platinum, the hydration level of the membrane, the concentration of the crossover reactants, and the temperature are the key factors influencing the membrane degradation rate. The model predicts that the low loading of small Pt particles in membrane causes fast degradation of the membrane, but the high loading of large Pt particles in the membrane mitigates membrane from chemical degradation in agreement with [9,19].